\documentclass[aps,prd,twocolumn,showpacs,superscriptaddress,showkeys]{revtex4-1}

\usepackage{amssymb}
\usepackage{amsmath}
\usepackage{amsfonts}
\usepackage{graphicx}
\usepackage{color}
\usepackage{xspace}
\usepackage{ulem}
\usepackage{mathtools}
\usepackage{hhline}

\newcommand{\AddrFCUPandCFP}{Departamento de F\'{\i}sica e Astronomia, Faculdade de Ci\^encias da Universidade do Porto and Centro de F\'{\i}sica do Porto, Rua do Campo Alegre 687, 4169-007, Porto, Portugal}

\newcommand{\AddrUC}{Departamento de F\'{\i}sica da Universidade de Coimbra and CFisUC \\ Rua Larga, 3004-516 Coimbra, Portugal}

\newcommand{\AddrCar}{Ottawa-Carleton Institute for Physics, Carleton University, 1125 Colonel By Drive, Ottawa, Ontario K1S 5B6, Canada}

\begin{document}


\title{Can dark matter drive electroweak symmetry breaking?}

\author{Catarina Cosme}    
\email{ccosme@physics.carleton.ca}
\affiliation{\AddrCar}
\affiliation{\AddrFCUPandCFP}

\author{Jo\~{a}o G.~Rosa} \email{jgrosa@uc.pt}\affiliation{\AddrUC}

\author{O. Bertolami}    \email{orfeu.bertolami@fc.up.pt}\affiliation{\AddrFCUPandCFP}
 
\date{\today}

\begin{abstract}

We consider the possibility of an oscillating scalar field accounting for dark matter and dynamically controlling the spontaneous breaking of the electroweak symmetry through a Higgs-portal coupling. This requires a late decay of the inflaton field, such that thermal effects do not restore the electroweak symmetry after reheating, and so inflation is followed by an inflaton matter-dominated epoch. During inflation, the dark scalar field acquires a large expectation value due to a negative nonminimal coupling to curvature, thus stabilizing the Higgs field by holding it at the origin. After inflation, the dark scalar oscillates in a quartic potential, behaving as dark radiation, and only when its amplitude drops below a critical value does the Higgs field acquire a non-zero vacuum expectation value. The dark scalar then becomes massive and starts behaving as cold dark matter until the present day. We further show that consistent scenarios require dark scalar masses in the few GeV range, which may be probed with future collider experiments.

\end{abstract}

\keywords{Dark matter, Scalar Field, Higgs boson, Electroweak symmetry breaking}


\maketitle




\section{Introduction}

The discovery of the Higgs boson at the Large Hadron Collider (LHC) opened up new windows to study the origin and nature of dark matter.
In fact, the study of interactions between dark matter and the Higgs boson has been increasing in interest in the literature, encompassing numerous
dark matter models, ranging from thermal candidates \cite{Patt:2006fw,Bento:2000ah,Bento:2001yk,MarchRussell:2008yu,Biswas:2011td,Pospelov:2011yp,Mahajan:2012nc,Cline:2013gha,Enqvist:2014zqa,Kouvaris:2014uoa,Costa:2014qga,Duerr:2015aka,Han:2015dua,Han:2015hda} to nonthermal ones \cite{Nurmi:2015ema,Tenkanen:2015nxa,Kainulainen:2016vzv,Bertolami:2016ywc,Cosme:2017cxk,Cosme:2018nly,Bernal:2018ins,Bernal:2018kcw,Alonso-Alvarez:2018tus,Alonso-Alvarez:2018ahf,Markkanen:2018gcw}.

In addition, the introduction of a dark scalar singlet may solve the Higgs vacuum stability problem. The Higgs vacuum is stable if its self-coupling, $\lambda_{h}$, is positive for any scale of energy $\mu$ where the minimum of its potential is a global minimum. However, for the measured Higgs mass $m_{h}\simeq125$
GeV, $\lambda_{h}$ becomes negative for energy scales around $\mu\sim10^{10}-10^{12}$ GeV  \cite{Degrassi:2012ry,Buttazzo:2013uya}, which are well below
the GUT or the Planck scales. This could constitute a problem since it may lead to a possible instability in the Higgs potential (see, for e.g., Refs. \cite{EliasMiro:2011aa, Degrassi:2012ry, Espinosa:2013lma} and references therein). The behavior of $\lambda_{h}$ is mostly driven by the large contribution of the top Yukawa coupling at one-loop, i.e., strongly depends on the top quark mass. When the coupling constant becomes negative, the renormalization group-improved Higgs potential is $V\left(h\right)=\lambda_{h}\frac{h^{4}}{4}<0$, and, therefore, the Higgs minimum could be only a local minimum, instead of a global minimum. However, if the time scale for quantum tunneling to this true minimum exceeds the age of the Universe, the Higgs vacuum is only metastable (see, for instance, Ref. \cite{Markkanen:2018pdo} for a review). In fact, Ref. \cite{Espinosa:2013lma} showed that the lifetime for quantum tunneling is much larger than the age of the Universe.

There have been several attempts to cure the (in)stability problem of the electroweak vacuum. For instance, Ref. \cite{Degrassi:2012ry} showed that
a shift in the top quark mass of about $\delta m_{t}=-2$ GeV would suffice to keep $\lambda_{h}>0$ at the Planck scale (this could also
be a good reason to motivate more precise measurements of the top quark mass). Other ways include introducing physics beyond the Standard
Model. In particular, coupling a scalar singlet with non-zero expectation value to the Higgs may stabilize the electroweak vacuum, provided
that the contribution of the coupling between the Higgs and the singlet scalar maintains the Higgs self-coupling positive. This idea has been
explored in the literature, and some of them promote this singlet scalar to a dark matter candidate, such as illustrated in Refs. \cite{Gonderinger:2009jp, Lebedev:2012zw, Cline:2013gha}. In addition, one must consider the stability of the Higgs field during inflation, since de Sitter quantum fluctuations could drive the field to the true global minimum of the potential. This may potentially be avoided if the Higgs field is sufficiently heavy during inflation, which may be achieved by coupling it to other fields such as the inflaton itself \cite{Lebedev:2012sy} or a dark matter scalar as we propose in this work.

Another interesting possibility, which we explore in this work, is
the case where a dark matter scalar coupled to the Higgs leads a nonthermal electroweak phase transition. We show that, besides stabilizing the Higgs field during inflation, a dark scalar may also completely alter the dynamics of the electroweak phase transition and we describe the conditions that must be satisfied for this to occur. As we will discuss, the electroweak symmetry may remain restored until very late even if the Universe's temperature is always below the weak scale. This could, in particular, preclude electroweak baryogenesis and the generation of gravitational waves through a thermal first-order phase transition.

We consider a self-interacting dark scalar field, $\Phi$, coupled to the Higgs field, $\mathcal{H}$, through a standard biquadratic ``Higgs-portal" coupling, and nonminimally coupled to gravity:
\begin{equation}
-\mathcal{L}_{int}=g^{2}\left|\Phi\right|^{2}\left|\mathcal{H}\right|^{2}+\lambda_{\phi}\left|\Phi\right|^{4}+V\left(\mathcal{H}\right)-\xi R\left|\Phi\right|^{2}~,\label{lagrangian EWSB}
\end{equation}
where the Higgs potential $V(\mathcal{H})$ takes the standard ``mexican hat" shape. For simplicity, we will take the nonminimal gravitational coupling of the Higgs field, which is allowed by the symmetries of the model, to be arbitrarily small so that we can neglect its effects. We note that including a positive Higgs nonminimal coupling to curvature would not change the qualitative behavior of the cosmological dynamics, namely the electroweak phase transition, and that this would, in fact, help stabilizing the Higgs during inflation in addition to the dark scalar as we discuss below  (see, for instance, Refs. \cite{Herranen:2014cua,Kamada:2014ufa,Espinosa:2015qea,Kohri:2016qqv,Kawasaki:2016ijp,Joti:2017fwe,Calmet:2017hja,Markkanen:2018pdo}).

 As in previous works \cite{Cosme:2017cxk, Cosme:2018nly}, we assume an underlying scale invariance of the theory, spontaneously broken by some mechanism that generates the Planck and electroweak mass scales in the Lagrangian, but which forbids a bare mass term for the dark scalar \cite{footnote0}.  It is thus easy to see that, for a sufficiently large value of $\Phi$, the minimum of the Higgs potential will lie at the origin, and it is natural to enquire whether the dark scalar can dynamically drive the spontaneous breaking of the electroweak symmetry \cite{footnote1}. While we assume that the dark scalar is only sensitive to the spontaneous breaking of scale invariance through its interactions with the Higgs field, there could be, of course, other sources and even mass terms explicitly breaking this symmetry. Our analysis will nevertheless hold if such contributions to the dark scalar's mass are subdominant compared to the contribution of the Higgs field, which is the regime on which we will focus in this work.

To prevent thermal effects from restoring the electroweak symmetry after inflation, we focus on scenarios with a late inflaton decay, such that the reheating temperature, $T_R$, is below $\sim100$ GeV. Consequently, inflation is followed by a long matter-dominated epoch while the inflaton oscillates about the origin in an approximately quadratic potential. As we will see in more detail below, the negative sign of the nonminimal coupling to gravity leads to a large expectation value for the dark scalar during inflation, which makes the Higgs field heavy and stabilizes it at the origin during this period. After inflation the dark scalar starts oscillating about the origin in its quartic potential, and its amplitude decreases with expansion, such that at some point it falls below a critical value that allows the Higgs to develop a non-zero vacuum expectation value. The spontaneous breaking of the electroweak symmetry is thus dynamically controlled by the dark scalar, and once it occurs the latter gains a mass and starts behaving as cold (pressureless) dark matter.

This work is organized as follows. In the next section we discuss the dynamics of the dark scalar and the Higgs field during inflation. In section 3 we describe the postinflationary dynamics of both fields, discussing the possibilities of reheating occurring before or after the electroweak symmetry is spontaneously broken. We discuss the consistency of our analysis and parametric constraints in section 4 and present our results for the allowed values of the dark scalar mass and couplings in section 5. We summarize our discussion and main conclusions in section 6.


\section{Inflation}

During inflation, the relevant interaction Lagrangian for the dynamics of the Higgs and dark scalar field, assuming they have no significant interactions with the inflaton field, is given by:
\begin{equation}
-\mathcal{L}_{inf}=\frac{g^{2}}{4}\,\phi^{2}h^{2}+\frac{\lambda_{\phi}}{4}\phi^{4}-\frac{\xi}{2}R\phi^{2}~,\label{relevant potential inf}
\end{equation}
where $\Phi=\frac{\phi}{\sqrt{2}}$, $\mathcal{H}=\frac{h}{\sqrt{2}}$ and the Ricci
scalar can be written in terms of the Hubble parameter,
\begin{equation}
R_{inf}\simeq12\,H_{inf}^2~,\label{Ricci inf}
\end{equation}
where $H_{inf}$ can be related to the tensor-to-scalar ratio of primordial curvature perturbations as:
\begin{equation}
H_{inf} \simeq 2.5\times 10^{13}\left({r\over 0.01}\right)^{1/2}\ \mathrm{GeV}~.
\label{Hinf}
\end{equation}
Since the interaction term between $\phi$ and $R$ has a negative
sign, the dark scalar acquires a vacuum expectation value (vev) during inflation, $\phi_{inf}$, with the minimum of the potential lying at:
\begin{equation}
\phi_{inf}=\sqrt{\frac{12\,\xi\,H_{inf}^{2}}{\lambda_{\phi}}},\qquad h_{inf}=0~.\label{phi inf complete}
\end{equation}
The dark scalar then provides a large mass to the Higgs field during inflation:
\begin{equation}
m_{h}=\frac{1}{\sqrt{2}}\,g\,\phi_{inf}\simeq\frac{g}{\sqrt{\lambda_{\phi}}}\,\sqrt{6\,\xi}\,H_{inf}~.\label{Higgs mass}
\end{equation}
We will see later that $g/\sqrt{\lambda_{\phi}}\sim10^{2}$ if the dark scalar accounts for all dark matter, such that $m_h\gtrsim H_{inf}$ for $\xi\gtrsim 10^{-5}$. This large Higgs mass has two related effects. First, it induces an additional quadratic term in the Higgs potential, thus shifting the field value at which the potential becomes unbounded (i.e.~$\lambda_h<0$) towards values larger than $H_{inf}$, i.e.~above the $10^{10}-10^{12}$ GeV scale at which it becomes unbounded in the Standard Model \cite{Buttazzo:2013uya}. Second, it suppresses the Higgs de Sitter quantum fluctuations, which for a light Higgs ($m_h\lesssim H_{inf}$) would be $\sim H_{inf}/2\pi\sim 10^{12}$ GeV unless the tensor-to-scalar ratio is very suppressed. For a massive Higgs field, the field variance during inflation on superhorizon scales is given by \cite{Riotto:2002yw}:
\begin{equation}
\left\langle h^{2}\right\rangle \simeq\left(\frac{H_{inf}}{2\pi}\right)^{2}\,\frac{H_{inf}}{m_{h}}~,\label{quantum fluct massive}
\end{equation}
which, using Eq. (\ref{Higgs mass}), simplifies to
\begin{equation}
\left\langle h^{2}\right\rangle \simeq\left(\frac{H_{inf}}{2\pi}\right)^{2}\,\frac{\lambda_{\phi}^{1/2}}{g\,\sqrt{6\,\xi}}~,\label{quantum fluct massive subs}
\end{equation}
corresponding to an average fluctuation amplitude $\sqrt{\left\langle h^{2}\right\rangle }\lesssim10^{11}$ GeV for $r\lesssim10^{-2}$ and $\xi\gtrsim 0.1$. Thus, the coupling between the Higgs and the dark scalar can prevent the former from falling into the putative large field true minimum during inflation.

We note that the dark scalar is also heavy during inflation, such that its de Sitter quantum fluctuations, with an amplitude $\sqrt{\langle\delta \phi^2\rangle}\simeq 0.05\,\xi^{-1/4}\,H_{inf}$ \cite{Cosme:2017cxk, Cosme:2018nly}, have a negligible effect on its expectation value $\phi_{inf}\gtrsim H_{inf}$, the latter setting the initial amplitude for field oscillations in the postinflationary epoch.


\section{Postinflationary period}

In this model we assume that, after inflation, the inflaton field,
$\chi$, does not decay immediately. Instead, the inflaton evolves
as nonrelativistic matter, while oscillating about the minimum of its potential, and an early matter era follows inflation until reheating finally occurs. Therefore, there are some significant changes in the dynamics of the Universe with respect to the usual radiation-dominated epoch. The scale factor evolves in time as $a\sim t^{2/3}$ and the Ricci scalar has a nonvanishing value, $R=3\,H^{2}$,
unlike its value during the radiation era ($R=0$). The evolution
of the inflaton energy density is thus given by:
\begin{equation}
\rho_{\chi}=3\,H_{end}^{2}\,M_P^{2}\,\left(\frac{a}{a_{end}}\right)^{-3}~,\label{inflaton evolution}
\end{equation}
where $M_P\simeq 2.4\times 10^{18}$ GeV is the reduced Planck mass and the subscript ``end" corresponds to the end of inflation. Note that $H_{end}$ depends on the particular inflationary model. Let us consider, for instance, the case where inflation is driven by a field with a power-law potential, $V\left(\chi\right)=\lambda\,\chi^{n}$. The number of e-folds of inflation, after the observable CMB scales become superhorizon, is given by:
\begin{equation}
N_{e}=-\frac{1}{M_P^{2}}\int_{\chi_{*}}^{\chi_{end}}\,\frac{V\left(\chi\right)}{V'\left(\chi\right)}\,d\chi\simeq{1\over2\,n}\frac{\chi_*^2}{M_P^{2}}~,\label{number efolds comp}
\end{equation}
where $\chi_{*}$ is the value of the inflaton field when observable CMB scales become superhorizon during inflation, with $\chi_{*}\gg\chi_{end}$. Inflation ends when $\epsilon= M_P^2 (V'/V)^2/2\sim 1$, yielding $\chi_{end}\simeq {n\over\sqrt{2}} M_P$, from which we deduce that:
\begin{equation}
H_{end}\simeq\left(\frac{\sqrt{n}}{2}\,\frac{1}{\sqrt{N_{e}}}\right)^{n/2}H_{inf}~.\label{Hend}
\end{equation}
According to Planck data, power-law potentials with $n\geq2$ are strongly disfavored \cite{Akrami:2018odb}, while models with e.g.~$n=2/3$ are compatible with the data. We will consider the above relation with $N_e=60$ henceforth in our discussion, using $H_{end}\sim0.4\,H_{inf}$ (corresponding to $n=2/3$) as a reference value, noting that for most models $H_{end}/H_{inf}\sim \mathcal{O}(0.1)$. Note that this model dependence is nevertheless degenerate with the unknown value of the tensor-to-scalar ratio, which we take as a free parameter.

At some stage, the inflaton decay reheats the Universe, establishing the beginning of the radiation-dominated epoch. This scenario resembles the so-called Polonyi problem found in many supergravity models, where the Polonyi field or other moduli decay at late times (see, for e.g., Refs. \cite{Coughlan:1983ci,Bertolami:1987xb,Bertolami:1988gz}). We assume that the inflaton transfers all its energy density into Standard Model degrees of freedom at a reheating temperature $T_{R}$:
\begin{equation}
\rho_{\chi}\left(a_{R}\right)=\frac{\pi^{2}}{30}\,g_{*R}\,T_{R}^{4}~,\label{inflaton reheating}
\end{equation}
where $g_{*R}$ is the number of relativistic degrees of freedom at reheating. The reheating temperature must be above $\sim10$ MeV, as the Universe must be radiation-dominated during Big Bang nucleosynthesis (BBN). As mentioned earlier, we will consider the case where reheating does not restore the electroweak symmetry, such that electroweak symmetry breaking is controlled by the dynamics of the dark matter scalar field, i.e,  $T_{R}\lesssim m_W\simeq 80$ GeV. Using Eqs. (\ref{inflaton evolution}) and (\ref{inflaton reheating}), the number of e-folds from inflation until reheating, $N_{R}$, 
reads:
\begin{align}
N_{R} & =-\frac{1}{3}\,\ln\left(\frac{\pi^{2}}{90}\,g_{*R}\,\frac{T_{R}^{4}}{M_P^{2}}\,\frac{1}{H_{end}^{2}}\right)\nonumber \\
 & \simeq46-\frac{1}{3}\ln\left(g_{*R}\right)-\frac{4}{3}\ln\left(\frac{T_{R}}{\mathrm{10\,GeV}}\right)\nonumber \\
 & +\frac{1}{3}\ln\left(\frac{r}{0.01}\right)+\frac{2}{3}\ln\left(\frac{H_{end}/H_{inf}}{0.4}\right)~.\label{N rh}
\end{align}
The interesting feature of this model is that the dark scalar will
control a nonthermal electroweak symmetry breaking (EWSB). From Eq.
(\ref{relevant potential inf}), it is easy to see that the minimum
of the Higgs potential occurs at 
\begin{equation}
\left|h\right|=\sqrt{\mathrm{v^{2}}-\frac{g^{2}\,\phi^{2}}{2\,\lambda_{h}}}~.\label{higgs min EWPT}
\end{equation}
EWSB then takes place when the amplitude of the field becomes smaller than the critical value:
\begin{equation}
\phi_{c}=\sqrt{2\,\lambda_{h}}\,\frac{\mathrm{v}}{g}~,\label{phi EW}
\end{equation}
noting that, in a few e-folds, the Higgs field should attain its final vacuum expectation value $\left|h\right|=\mathrm{v}$. This assumes that the temperature of the Universe when the field reaches the critical value is already below the electroweak scale, since otherwise the Higgs field would remain in the symmetric phase until much later and one would have the conventional thermal electroweak phase transition. In scenarios where the inflaton has a constant decay width, the maximum temperature attained during the reheating process is typically much larger than the final reheating temperature when radiation becomes dominant, which could be in tension with the above assumption. However, one can envisage scenarios where the temperature of the Universe never exceeds (significantly) the reheating temperature, such as in e.g. Refs. \cite{Bastero-Gil:2015lga,Manso:2018cba} where inflaton decay is kinematically blocked for most of its oscillation period. In these models the inflaton is directly coupled only to two fermion fields, for instance right-handed neutrinos or milli-charged particles, with masses $m_\pm= |m_f\pm h\chi|$ as imposed by a discrete interchange symmetry, where $h$ denotes the Yukawa coupling. For $m_f> m_\chi/2$, inflaton decays occur only within a narrow field range, i.e.~a small fraction of each oscillation period, making the decay quite inefficient. These fermions may either decay or annihilate into Standard Model states. As can be seen in Fig.~2 of \cite{Bastero-Gil:2015lga}, the radiation energy density, and, hence, also the temperature, remains approximately constant until it takes over the inflaton, and this temperature can be arbitrarily small if the Yukawa coupling is sufficiently suppressed (c.f.~Eq.~(2.13) of Ref.~\cite{Bastero-Gil:2015lga}). In this parametric regime, the inflaton, albeit stable at late times, will only give a subdominant contribution to the present dark matter abundance. Higgs-inflaton couplings are not needed for reheating in this case and may, in fact, be extremely suppressed in a technically natural way depending on how the fermions are coupled to Standard Model states, so in this framework there are certainly scenarios where Higgs production from the inflaton field can be neglected, as we will assume henceforth.

 In the remainder of our discussion, we will therefore assume a reheating process similar to the one described in Refs. \cite{Bastero-Gil:2015lga,Manso:2018cba}, such that it suffices to require $T_R\lesssim 80$ GeV to ensure that the electroweak phase transition is driven by the dark scalar field.

In the following subsections, we will study the dynamics of the dark scalar when reheating occurs after or before EWSB. Note, however, that $N_{R}$ is determined solely by $r$ and $T_{R}$, being independent of when EWSB takes place. Hence, our model has five free parameters: $r$, $\xi$, $g$, $\lambda_{\phi}$, and $T_{R}$. Note that there is also a mild dependence on the inflationary model through the ratio $H_{end}/H_{inf}\sim\mathcal{O}(0.1)$.


\subsection{Reheating after EWSB}

\begin{figure}[htbp]
\begin{centering}
\includegraphics[scale=0.365]{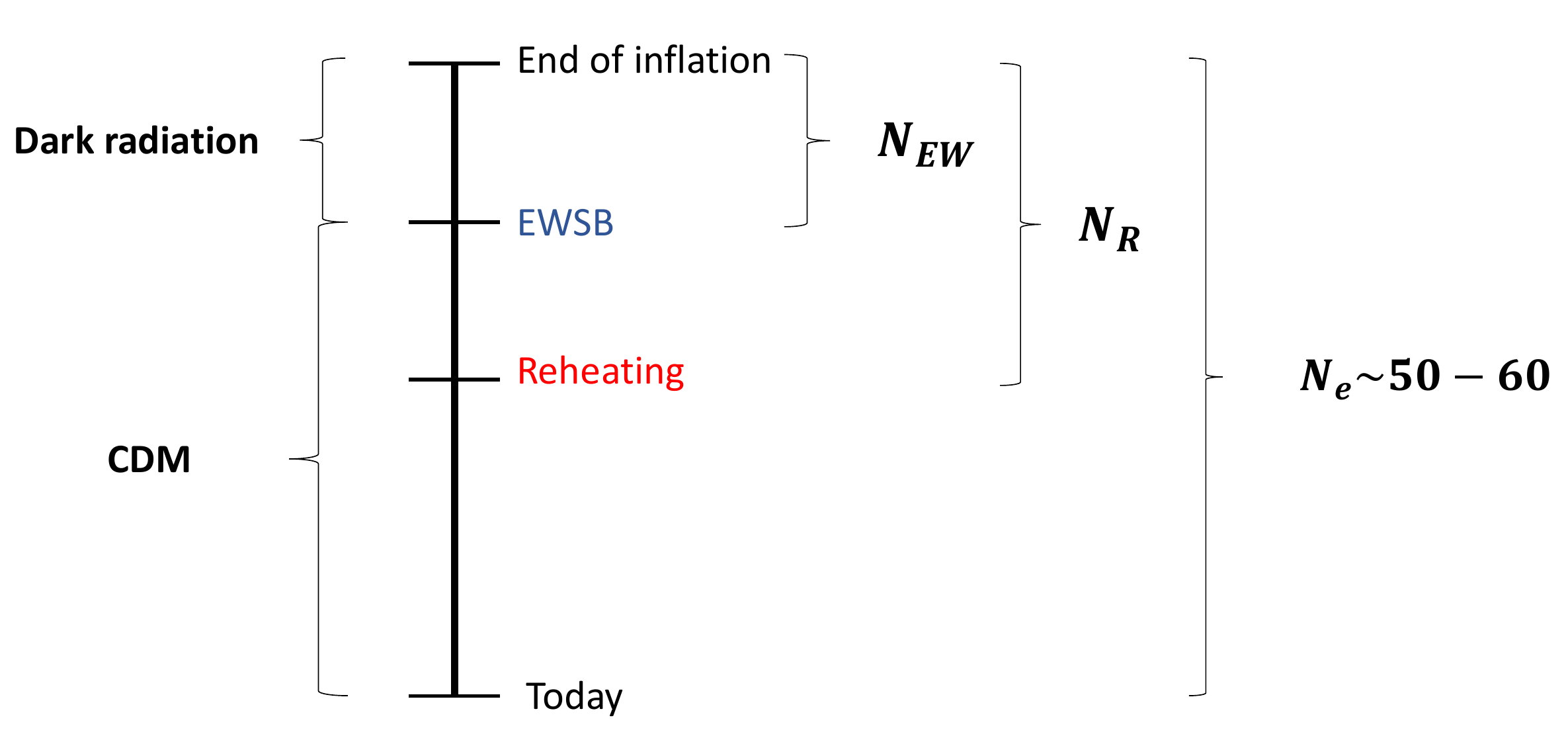}\vspace{-0.2cm}
\par\end{centering}
\caption{Time scale of the events: in this scenario, reheating occurs after
EWSB. The dark scalar behaves like dark radiation until EWSB and like
CDM afterwards. $N_{R}$ corresponds to the number of e-folds from
inflation until reheating and $N_{EW}$ is the number of
e-folds from inflation until EWSB. \label{fig:Reheating-after-the}}
\end{figure}

The first scenario we study is the one where reheating occurs
after EWSB, as illustrated in Fig. \ref{fig:Reheating-after-the}.

Before EWSB, the quartic term dominates the energy density of the dark scalar and it starts oscillating about the origin with initial amplitude $\phi_{inf}$. The amplitude decays as $\phi\propto a^{-1}$, such that the field behaves as dark radiation, $\rho_\phi \propto a^{-4}$. Note that $R\propto H^2\propto a^{-3}$, so that the effects of the nonminimal coupling to gravity decay faster than those of the quartic self-interactions and may thus be neglected. We will assume, for simplicity, that once the electroweak symmetry is spontaneously broken and the field becomes massive the associated quadratic term in the scalar potential becomes dominant, such that the field behaves as cold dark matter (CDM) from EWSB onwards. Therefore, the dark scalar exhibits two behaviors:
\begin{equation}
\begin{cases}
\phi_{rad}\left(a\right)=\phi_{inf}\,\left(\frac{a}{a_{end}}\right)^{-1}, & a_{end}<a<a_{c}\\
\phi_{DM}\left(a\right)=\phi_{c}\,\left(\frac{a}{a_{c}}\right)^{-3/2}, & a>a_{c}
\end{cases}~,\label{phi behaviours}
\end{equation}
where $a_{c}$ is the value of the scale factor at which EWSB takes place. At EWSB, we have: 
\begin{equation}
\phi_{c}=\phi_{inf}\,\left(\frac{a_{c}}{a_{end}}\right)^{-1}\label{EWPT phi inf phi EW}
\end{equation}
and, therefore, the number of e-folds from inflation until EWSB, $N_{EW}$, is given by
\begin{align}
N_{EW} & =\ln\left(\phi_{inf}/\phi_c\right)\simeq 27+\ln\left(\sqrt{\xi}\,\frac{g}{\sqrt{\lambda_{\phi}}}\,\left(\frac{r}{0.01}\right)^{1/2}\right)~.\label{New 1}
\end{align}
Once reheating occurs, the Universe enters the usual
radiation-dominated epoch. Thus, the number of dark matter particles in a comoving volume, $n_{\phi}/s$,
becomes constant. The dark scalar amplitude at reheating is, then:
\begin{align}
\phi_{R} & \equiv\phi\left(a_{R}\right) =\phi_{c}\,\left(\frac{a_{R}}{a_{c}}\right)^{-3/2} =\phi_{c}\,e^{-\frac{3}{2}\,\left(N_{R}-N_{EW}\right)}~,\label{phi arh}
\end{align}
where 
%

\begin{align}
N_{R}-N_{EW} & =19-\frac{1}{3}\,\ln\left(g_{*R}\right)-\frac{4}{3}\ln\left(\frac{T_{R}}{\mathrm{10\,GeV}}\right)\nonumber \\
 & -\ln\left(\sqrt{\xi}\,\frac{g}{\sqrt{\lambda_{\phi}}}\,\left(\frac{r}{0.01}\right)^{\frac{1}{6}}\right)\nonumber \\
 & +~\frac{2}{3}\ln\left(\frac{H_{end}/H_{inf}}{0.4}\right)~\label{New2}
\end{align}
Introducing the last equation into Eq. (\ref{phi arh}), the amplitude
of the field at reheating becomes: 
\begin{align}
\phi_{R} & \simeq10^{-10}\,\sqrt{g_{*R}}\,\left(\frac{r}{0.01}\right)^{1/4}\,\left(\frac{T_{R}}{\mathrm{10\,GeV}}\right)^{2}\nonumber \\
 & \times\xi^{3/4}g^{1/2}\lambda_{\phi}^{-3/4}\left(\frac{H_{end}/H_{inf}}{0.4}\right)^{-1}\ \mathrm{GeV.}~\label{phi arh num}
\end{align}
The number of particles in a comoving volume at $T_{R}$ is, then:
\begin{equation}
\left(\frac{n_{\phi}}{s}\right)_{R}={45\over 4\pi^2} {m_\phi\phi_R^2\over g_{*R}T_R^3}~,\label{n s rh}
\end{equation}
where $m_{\phi}$ stands for the dark scalar mass once the electroweak symmetry is spontaneously broken and is given by:
\begin{equation}
m_{\phi}=\frac{1}{\sqrt{2}}\,g\,\mathrm{v}~.\label{phi mass}
\end{equation}
The present dark matter abundance then reads:
\begin{align}
\Omega_{\phi,0} & =\frac{m_{\phi}}{3\,H_{0}^{2}\,M_P^{2}}\left(\frac{n_{\phi}}{s}\right)_{R}\,s_{0}\nonumber \\
 & =\frac{m_{\phi}^{2}}{6\,H_{0}^{2}\,M_P^{2}}\,\phi_{R}^{2}\,\frac{g_{*0}}{g_{*R}}\,\left(\frac{T_{0}}{T_{R}}\right)^{3}~,\label{omega today}
\end{align}
where $g_{*0}$, $T_{0}$ and $H_{0}$ are the present values of the
number of relativistic degrees of freedom, CMB temperature and Hubble
parameter, respectively.

Replacing Eq. (\ref{phi arh num}) into the last expression and fixing
$\Omega_{\phi,0}=0.26$, we then obtain a relation between $g$ and $\lambda_{\phi}$: 
\begin{align}
g & \simeq9\times10^{2}\,\left(\frac{T_{R}}{10\,\mathrm{GeV}}\right)^{-1/3}\,\left(\frac{r}{0.01}\right)^{-1/6}\nonumber \\
 & \times \xi^{-1/2}\left(\frac{H_{end}/H_{inf}}{\mathrm{0.4}}\right)^{2/3}\,\lambda_{\phi}^{1/2}~.\label{relation between g and lambda}
\end{align}
%


\subsection{Reheating before EWSB}

\begin{figure}[htbp]
\begin{centering}
\includegraphics[scale=0.365]{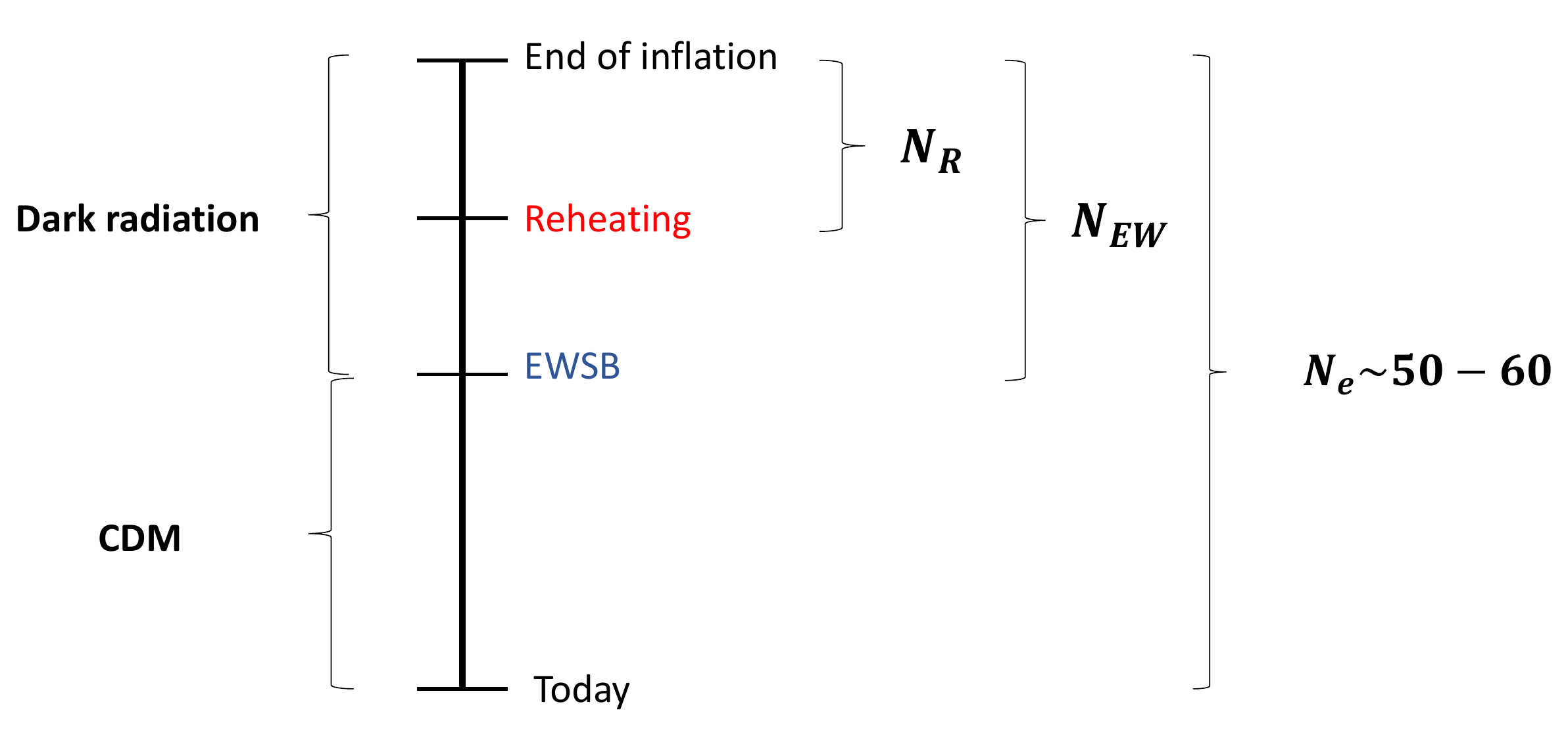}\vspace{-0.2cm}
\par\end{centering}
\caption{Time scale of the events: in this putative scenario, reheating occurs before
EWSB. The dark scalar behaves like dark radiation until EWSB and like
CDM afterwards. $N_{R}$ corresponds to the number of e-folds from
inflation until reheating and $N_{EW}$ is the number of
e-folds from inflation until EWSB.\label{fig:Reheating-before-the-}}
\end{figure}

The second putative scenario we should consider is the case where reheating occurs before EWSB, as illustrated in Fig. \ref{fig:Reheating-before-the-}.

Since the number of e-folds from inflation until reheating does not depend
on when EWSB takes place, $N_{R}$ is given by Eq. (\ref{N rh}) as in the previously discussed scenario.
Similarly, $N_{EW}$ only depends on $\phi_{inf}$ and $\phi_{c}$
and, therefore, it is given by Eq. (\ref{New 1}). The
difference between this and the previous scenario is that $N_{EW}$ should now exceed $N_{R}$. The dark scalar behaves like dark radiation
from reheating until EWSB, after which $n_{\phi}/s$ becomes constant. From reheating onwards, the Universe enters the usual radiation-dominated epoch and $R=0$.

The amplitude of the field at reheating is different from the
previous scenario:
\begin{equation}
\phi_{R}=\phi_{inf}\,\left(\frac{a_{R}}{a_{end}}\right)^{-1} =\phi_{inf}\,e^{-N_{R}}~,\label{a rh before}
\end{equation}
and now we have a defined temperature and can write the amplitude
of the field as a function of the temperature: 
\begin{equation}
\phi_{rad}\left(T\right)=\phi_{R}\,\frac{T}{T_{R}}\,\left(\frac{g_{*T}}{g_{*R}}\right)^{1/3}~.\label{phi rad after reheat}
\end{equation}
This can be used to compute the temperature at which EWSB occurs,
$T_{c}$:
\begin{equation}
T_{c}=\frac{\phi_{c}}{\phi_{R}}\,T_{R}\,\left(\frac{g_{*c}}{g_{*R}}\right)^{-1/3}~.\label{TEW}
\end{equation}
At $T_{c}$ the dark scalar stops holding the Higgs at the origin. Notice, however, that $T_{c}$ must be smaller than the usual $T_{EW}\sim80$ GeV,
so that the dark scalar can control the EWSB and the latter is not restored by thermal effects. By proceeding as in the previous subsection, since $n_{\phi}/s$ is constant as soon the field starts behaving as CDM, the present
dark matter abundance is given by:
\begin{equation}
\Omega_{\phi,0}=\frac{m_{\phi}^{2}}{6\,H_{0}^{2}\,M_P^{2}}\,\phi_{c}^{2}\,\frac{g_{*0}}{g_{*c}}\,\left(\frac{T_{0}}{T_{c}}\right)^{3}.\label{today's abundance before}
\end{equation}
Setting $\Omega_{\phi,0}=0.26$ we then obtain for the temperature at which the field amplitude falls below the critical value:
\begin{align}
T_{c} & =\left(\frac{2\,\lambda_{h}\,\mathrm{v}^{4}}{12\,H_{0}^{2}\,M_P^{2}}\right)^{1/3}\,\left(\frac{g_{*0}}{g_{*c}}\right)^{1/3}\,\frac{T_{0}}{\Omega_{\phi,0}^{1/3}}\nonumber \\
 & \sim7\times10^{5}\,\left(\frac{g_{*0}}{g_{*c}}\right)^{1/3}\,\mathrm{GeV}~.\label{Tc}
\end{align}
Hence, we conclude that, for reheating to occur before EWSB, $T_c$ must be well above $T_{EW}\sim80$ GeV.  This is not consistent with our reasoning given that, at that temperature, the Higgs thermal mass is still sufficiently large to hold the latter at the origin, such that EWSB does not occur at $T_{c}$ as assumed and, consequently, the dark scalar remains massless and behaves as dark radiation, as opposed to our starting assumption. In the remainder of this paper, we will thus focus only on the case where reheating occurs after EWSB, given that in this scenario the dark scalar, in addition to being a viable dark matter candidate, can also control a nonthermal EWSB.

\section{Consistency analysis}
\label{sec: Model const}

In analyzing the dynamics of the dark scalar and of the Higgs field both during and after inflation we have made several technical assumptions. In this section, we discuss the parametric constraints imposed by these assumptions and also by the properties of the Higgs boson measured at the Large Hadron Collider (LHC).

First, our scenario assumes that inflation is driven by a scalar field, $\chi$, that is neither the dark scalar nor the Higgs field. Therefore,
we have to ensure, in particular, that the dark scalar does not affect the dynamics of inflation. The dark scalar's contribution to the effective potential during inflation is given by:
\begin{equation}
V\left(\phi_{inf}\right)  \simeq\frac{\lambda_{\phi}}{4}\phi_{inf}^{4}-\frac{\xi}{2}\,R\,\phi_{inf}^{2}
  \simeq-\frac{12^{2}}{4}\xi^{2}\,\frac{H_{inf}^{4}}{\lambda_{\phi}}~.\label{phi energy density infl}
\end{equation}
Requiring that this does not significantly reduce the inflationary energy density $V(\chi)\simeq 3 H_{\inf}^2 M_P^2$ then implies the condition:
\begin{equation}
\phi_{inf}<\frac{M_P}{\sqrt{\xi}}~,\label{condition: phi is not inflaton}
\end{equation}
which constrains the allowed values of the nonminimal coupling $\xi$ and the self-coupling $\lambda_\phi$, depending on the tensor-to-scalar ratio, i.e.~the scale of inflation:
\begin{equation}
\lambda_\phi > 12 \xi^2 {H_{inf}^2\over M_P^2} \simeq 1.3\times10^{-9}\xi^2\left({r\over 0.01}\right)~.
\end{equation}
Second, we have assumed that the dark scalar field starts behaving as CDM as soon as the electroweak symmetry is spontaneously broken, i.e.~when the field amplitude falls below the critical value $\phi_c$.  This means that the quadratic term has to dominate over the quartic term at EWSB, that is, $g^{2}\,\mathrm{v^{2}\,\phi_{c}^{2}}/\left(\lambda_{\phi}\,\phi_{c}^{4}\right)>1$,
which translates into the following condition:
\begin{align}
g & ^{4}>2\,\lambda_{h}\lambda_{\phi}\,.\label{condition: CDM at EWPT}
\end{align}
Finally, radiative corrections to the quartic coupling from the Higgs-portal coupling should be small, unless we accept some degree of fine-tuning:
\begin{equation}
\delta\lambda_{\phi}\sim\frac{g^{4}}{16\pi^{2}}<\lambda_{\phi}\,.\label{radiative corrections}
\end{equation}
From the experimental point of view, the Higgs may decay into dark scalar pairs with a decay width
\begin{equation}
\Gamma_{h\rightarrow\phi\phi}\simeq\frac{1}{8\pi}\,\frac{g^{4}\mathrm{v^{2}}}{4\,m_{h}},\label{decay width inv}
\end{equation}
leading to a Higgs branching ratio into invisible particles, assuming $\Gamma_{h\rightarrow inv}=\Gamma_{h\rightarrow\phi\phi}$ :
\begin{equation}
\ensuremath{\mathrm{Br}_{inv}=\frac{\Gamma_{h\rightarrow \phi\phi}}{\Gamma_{h}+\Gamma_{h\rightarrow \phi\phi}}}.\label{branching ratio}
\end{equation}
Current limits from the LHC establish an upper bound for the branching ratio $\mathrm{Br}_{inv}<0.23$ \cite{Aad:2015pla}, and using $\ensuremath{\Gamma_{h}=4.07\times10^{-3}}$ GeV \cite{Cuoco:2016jqt}, this yields an upper bound on the Higgs-portal coupling:
\begin{equation}
g<0.13~,\label{branch bound}
\end{equation}
which translates into an upper bound $m_\phi \lesssim 22.6$ GeV.

From the dynamical perspective, we have also implicitly assumed that the dark scalar field remains in the form of an oscillating condensate, such that processes that may lead to its evaporation and subsequent thermalization (which would yield a Weakly Interacting Massive Particle (WIMP)-like dark matter candidate) must be inefficient, as we discuss in detail below.


\subsection{Condensate evaporation}

The dark scalar provides mass to the Higgs field during the period before EWSB. Since $\phi$ is oscillating, this could induce oscillations of the Higgs mass. This may pose a problem, since if the Higgs mass $m_{h}<\sqrt{3\,\lambda_{\phi}}\,\phi_{rad}$, (perturbative or nonperturbative) Higgs production by the oscillating condensate is kinematically allowed  and can lead to the condensate's evaporation.

A solution to this problem is to provide initial conditions to the field such that its absolute value, and hence the Higgs mass, does not oscillate. This is possible if the dark scalar oscillates in the complex plane, as e.g.~in the Affleck-Dine (AD) mechanism for baryogenesis  \cite{Affleck:1984fy, Dine:1995kz}. We may then consider additional nonrenormalizable terms in the dark scalar's potential, particularly the so-called A-terms that explicitly break the underlying global U(1) symmetry. These are well-motivated within e.g.~supersymmetric theories, but more generally we expect such terms to be present since gravity in particular should not respect global symmetries. For instance, we may consider adding the following nonrenormalizable terms to the scalar potential:
\begin{equation}
V_{NR}(\phi)=\frac{a}{5M_P}\phi^5+\mathrm{h.c.}+\frac{b}{6M_P^2}\,\left|\phi\right|^{6}~, \label{AD Lag}
\end{equation}
where $a$ and $b$ are dimensionless couplings, the former taking complex values. We have chosen, as an illustrative case, the lowest dimension nonrenormalizable operators, with the $b$-term being required for the potential to be bounded from below, but higher-dimensional operators are also possible. Contributions of this form may arise from different sources, and as for the AD field the phase of the dimension-5 A-term may be different during and after inflation, thus yielding a different phase for the field at the minimum of the potential in these two periods. This, in turn, implies that the phase of the field will vary after inflation, thus generating ``angular momentum" in field space, $n_\phi= 2|\phi|^2\dot\theta$ for $\phi=|\phi|e^{i\theta}$. It is also possible that $a$ vanishes during inflation such that initially $\theta$ takes random values in different inflationary patches. The initial phase misalignment in our inflationary patch will then generate the required angular motion after inflation ends, which is known as spontaneous CP-violation. 

\begin{figure}[htbp]
\begin{centering}
\includegraphics[scale=0.55]{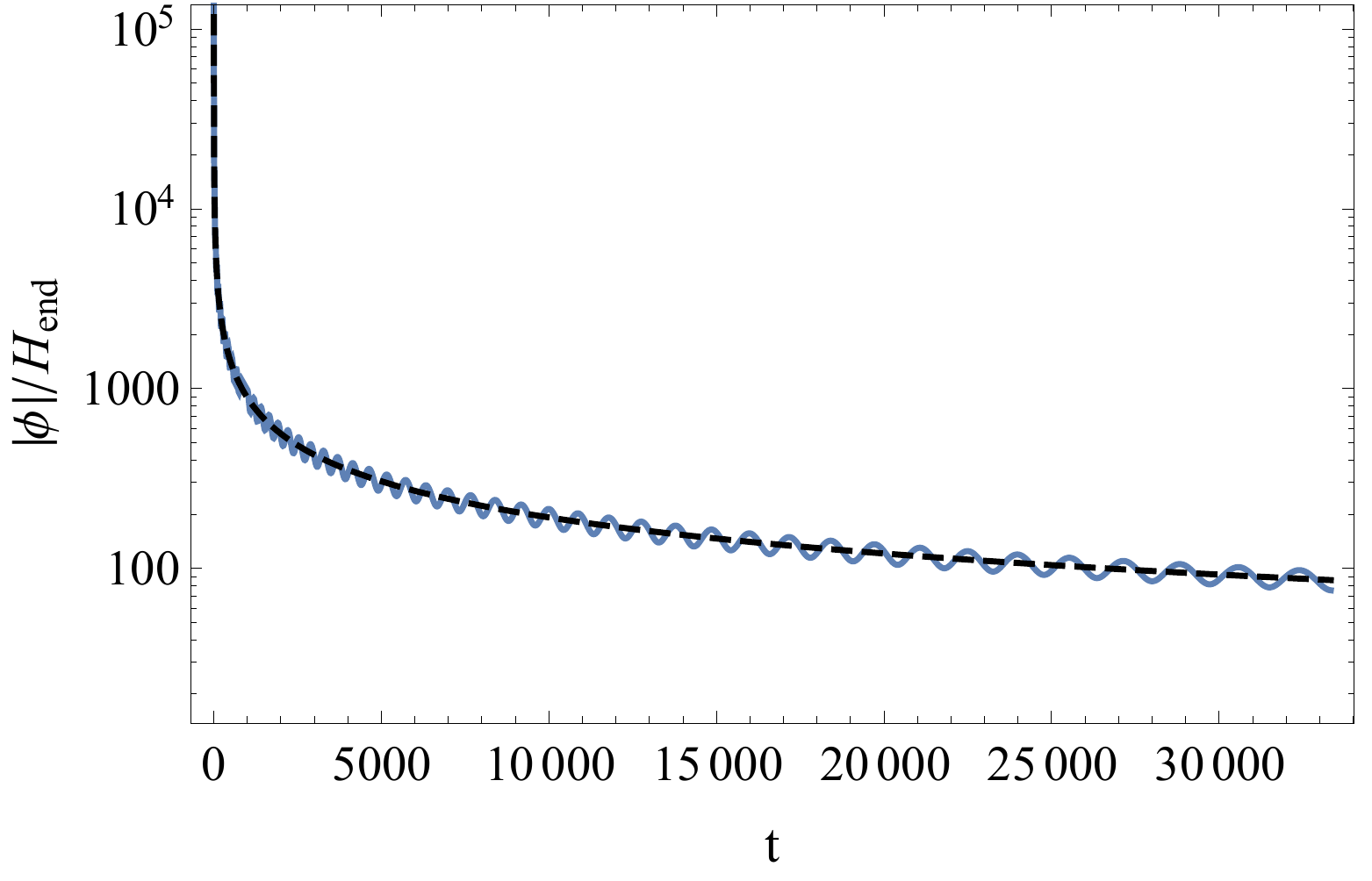}\vspace{0.5cm}
\includegraphics[scale=0.5]{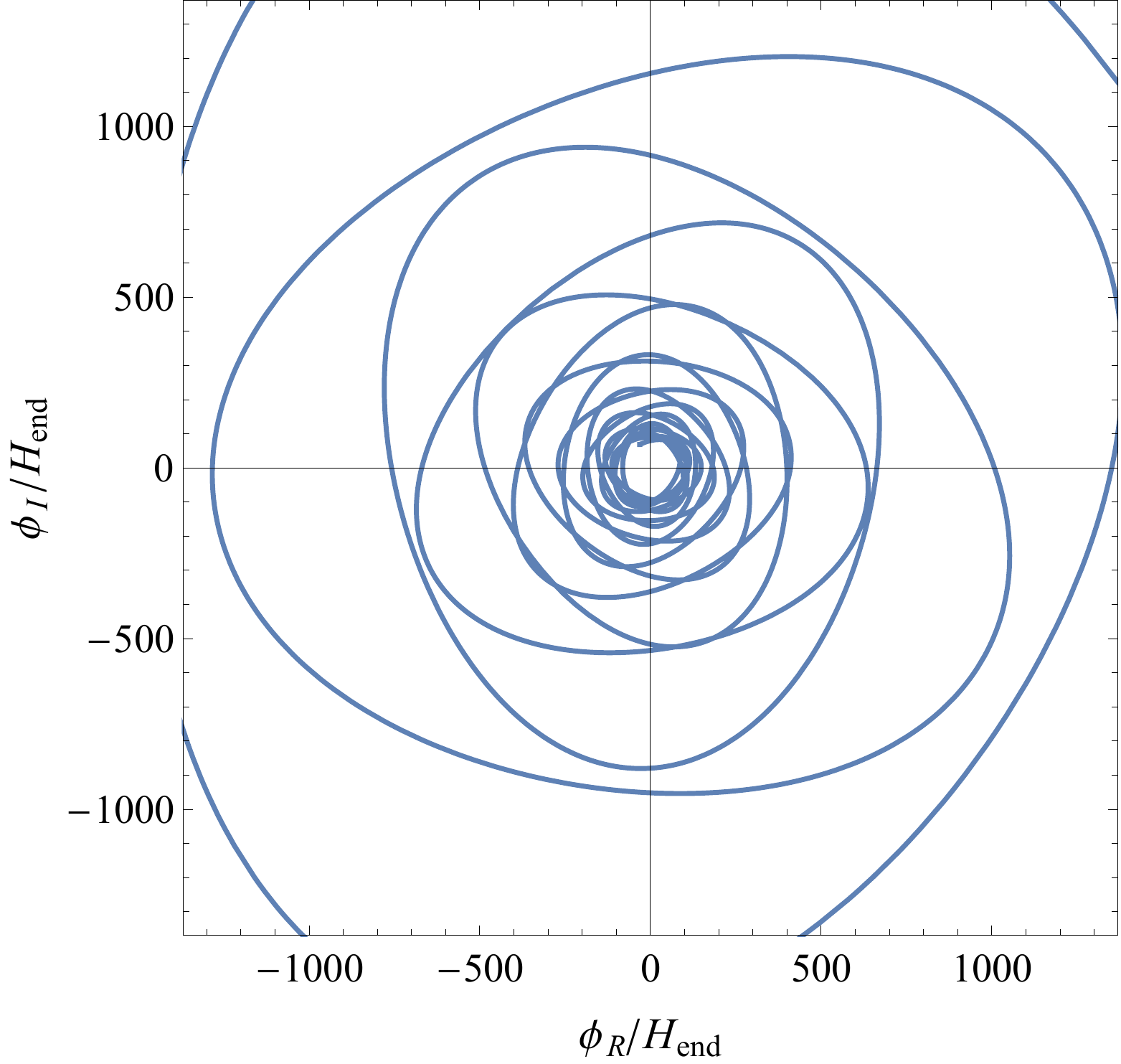}
\par\end{centering}
\caption{[Top] Evolution of $\left|\phi\right|/H_{end}$ as a function of time  (solid
blue curve), compared to the $\left|\phi\right|\sim t^{-2/3}\sim a^{-1}$ behavior (dashed black curve). [Bottom] Field evolution in the complex plane. We have considered $\xi=1$, $\lambda_{\phi}=10^{-9}$, $a\simeq 0.1 \sqrt{\lambda_\phi^3/\xi}M_P/H_{end}$, $b\simeq 0.1(\lambda_\phi^2/\xi)(M_P/H_{end})^2 $, and $\theta=1.5$ rad.}
\label{AD}
\end{figure}

To illustrate the motion of the field after inflation, we have numerically evolved the field equation in the postinflationary matter era for an example where $a$ is real and positive, and both $a$ and $b$ are chosen such that the effects of the nonrenormalizable terms are significant just after inflation, without changing considerably the magnitude of the field's expectation value acquired during inflation. According to the discussion above, we take the initial phase of the field in the postinflationary era to be away from the minima of the potential, which occur for $\theta_{n}=n\pi/5$ with odd $n$ in this case. The results are shown in Fig.~\ref{AD}.

As one can see in this figure, the field spirals in the complex plane after acquiring angular momentum due to the $A$-term. This means that $|\phi|$, and, hence, the Higgs mass, oscillates with a small amplitude around its mean value, and the Higgs never becomes light enough to be produced by the oscillations of the dark field. Both nonrenormalizable terms decay faster than the quartic term in the potential, such that on average $|\phi|\sim a^{-1}$ and the dark scalar behaves as dark radiation as we have assumed in our general discussion.

Since reheating can only consistently occur after EWSB as we have shown above, the only other possible channel for the evaporation of the dark scalar field is the perturbative production of $\phi$-particles by the oscillating background condensate. The dark scalar behaves like radiation until EWSB and the condensate decay width is given by \cite{Cosme:2017cxk, Cosme:2018nly}:
\begin{equation}
\Gamma_{\phi\rightarrow\delta\phi\delta\phi}\simeq4\times10^{-2}\,\lambda_{\phi}^{3/2}\,\phi_{rad},\label{condensate decay width}
\end{equation}
where, at EWSB, $\phi_{rad}=\phi_{c}$ . Condensate evaporation is then avoided if this never exceeds the Hubble expansion rate until EWSB, noting that after EWSB this production channel is blocked since the dark scalar becomes massive (see e.g.~\cite{Cosme:2017cxk, Cosme:2018nly}):
\begin{equation}
\left.\frac{\Gamma_{\phi\rightarrow\delta\phi\delta\phi}}{H}\right|_{c}<1~.\label{EW condition evaporation}
\end{equation}
Since the Universe is still in a matter-dominated era at EWSB, the Hubble parameter can be computed using
the expression for the inflaton's energy density given in Eq. (\ref{inflaton evolution}):
\begin{equation}
H_{c}^{2}=\frac{\rho_{\chi}\left(a_{c}\right)}{3\,M_P^{2}}=H_{end}^{2}\,\left(\frac{\phi_{c}}{\phi_{inf}}\right)^{3}.\label{H parameter EW reheating after}
\end{equation}
Therefore, from Eq. (\ref{EW condition evaporation}) we find that 
\begin{equation}
g<10^{-11}\,\left(\frac{r}{0.01}\right)^{-1/2}\,\xi^{-3/2}\,\lambda_{\phi}^{-3/2}\,\left(\frac{H_{end}/H_{inf}}{0.4}\right)^{2},\label{upper bound g reheating after}
\end{equation}
and using the relation between $g$ and $\lambda_{\phi}$ (Eq. (\ref{relation between g and lambda})),
the upper bound on $g$ reads
\begin{equation}
g<0.4\,\left(\frac{T_{R}}{10\,\mathrm{GeV}}\right)^{-1/4}\,\left(\frac{r}{0.01}\right)^{-1/4}\,\xi^{-3/4}\,\left(\frac{H_{end}/H_{inf}}{0.4}\right).\label{upper bound g final reheating after}
\end{equation}

In addition, $\phi$ particles may annihilate
into fermions, via virtual Higgs exchange, which might thermalize
the condensate after reheating. The interaction rate of this process
is given by:
\begin{equation}
\Gamma_{\phi\phi\rightarrow f\overline{f}}=n_{\phi}\left\langle \sigma v\right\rangle _{fermions},\label{production rate of the process}
\end{equation}
where $n_{\phi}=m_{\phi}\phi^{2}$ is the number density of $\phi$
particles and $\left\langle \sigma v\right\rangle _{fermions}$, the
cross section for dark matter annihilation into fermions that was
computed in Ref. \cite{Cosme:2018nly}, reads: 
\begin{align}
\left\langle \text{\ensuremath{\sigma}}v_{rel}\right\rangle _{fermions} & \simeq\frac{N_{c}}{2\pi}\,\frac{1}{\mathrm{v^{2}}}\,\frac{m_{\phi}^{4}}{\left(4m_{\phi}^{2}-m_{h}^{2}\right)^{2}+m_{h}^{2}\Gamma_{h}^{2}}\nonumber \\
 &\times \left(\frac{m_{f}}{\mathrm{v}}\right)^{2}\,\left(1-\frac{m_{f}^{2}}{m_{\phi}^{2}}\right)^{3/2}\label{cross section}
\end{align}
where $N_{c}$ is the number of colors ($N_{c}=1$ for leptons and
$N_{c}=3$ for quarks) and $\Gamma_{h}=4.07\times10^{-3}\,\mathrm{GeV}$
is the total Higgs decay rate. Combining Eqs. (\ref{phi arh num}) and (\ref{relation between g and lambda}), the amplitude
of the field at reheating is: 
\begin{equation}
\phi_{R}\simeq2.7\times10^{-6}\,\sqrt{g_{*R}}\,\left(\frac{T_{R}}{\mathrm{10\,GeV}}\right)^{3/2}\,g^{-1}\ \mathrm{GeV}.\label{rh with relation g lambda}
\end{equation}
while the Hubble parameter is:
\begin{equation}
H_{R}=\sqrt{\frac{\pi^{2}}{90}\,g_{*R}}\,\frac{T_{R}^{2}}{M_P}.\label{Hubble parameter reheating}
\end{equation}
Comparing the interaction rate $\Gamma_{\phi\phi\rightarrow f\overline{f}}$
with the Hubble parameter, we get:
\begin{equation}
\frac{\Gamma_{\phi\phi\rightarrow f\overline{f}}}{H_{R}}\simeq3\times10^{-9}\,\sqrt{g_{*R}}\,\left(\frac{T_{R}}{10\,\mathrm{GeV}}\right)\,\left(\frac{m_{f}}{\mathrm{GeV}}\right)^{2}\,\left(\frac{m_{\phi}}{\mathrm{GeV}}\right)^{3}\,N_{c},\label{annihilation over H}
\end{equation}
which means that the process is negligible at reheating for the range
of mass and temperature allowed in this model ($m_{\phi}\lesssim10$
Gev, $10\,\mathrm{MeV}<T_{R}<80$ GeV) and, therefore, it does not
lead to the thermalization of the condensate after reheating. There are other potentially significant scattering processes involving light fermions in the thermal bath, but we will not consider them in further detail since they do not change the number of $\phi$ particles per comoving volume, $n_{\phi}/s$, and hence the dark matter abundance.
%

\section{Results}

\begin{figure}[htbp]
\begin{centering}
  \begin{minipage}[b]{0.45\textwidth}
  \includegraphics[width=0.8\textwidth]{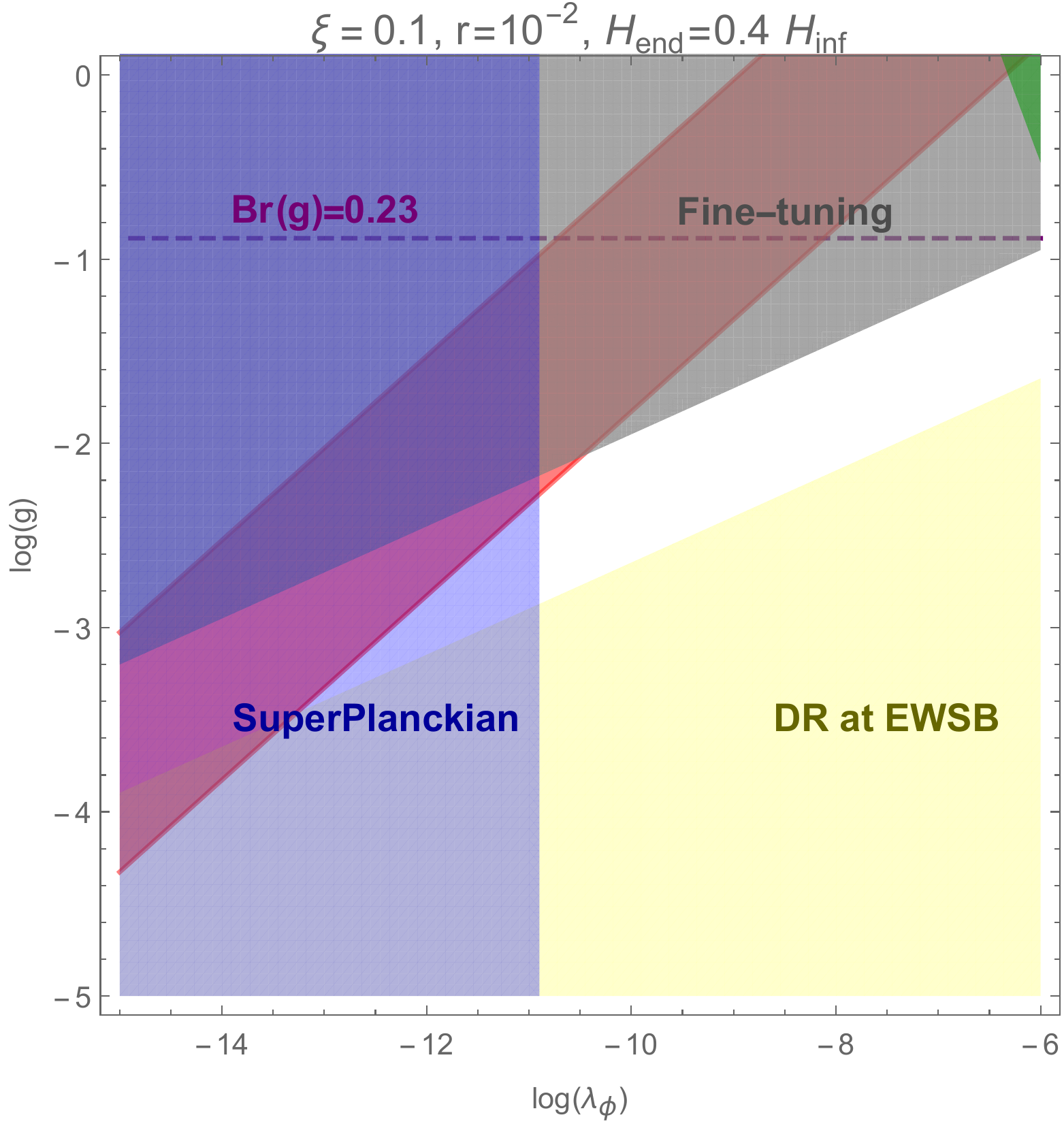}
  \end{minipage}
   \begin{minipage}[b]{0.45\textwidth}
  \includegraphics[width=0.8\textwidth]{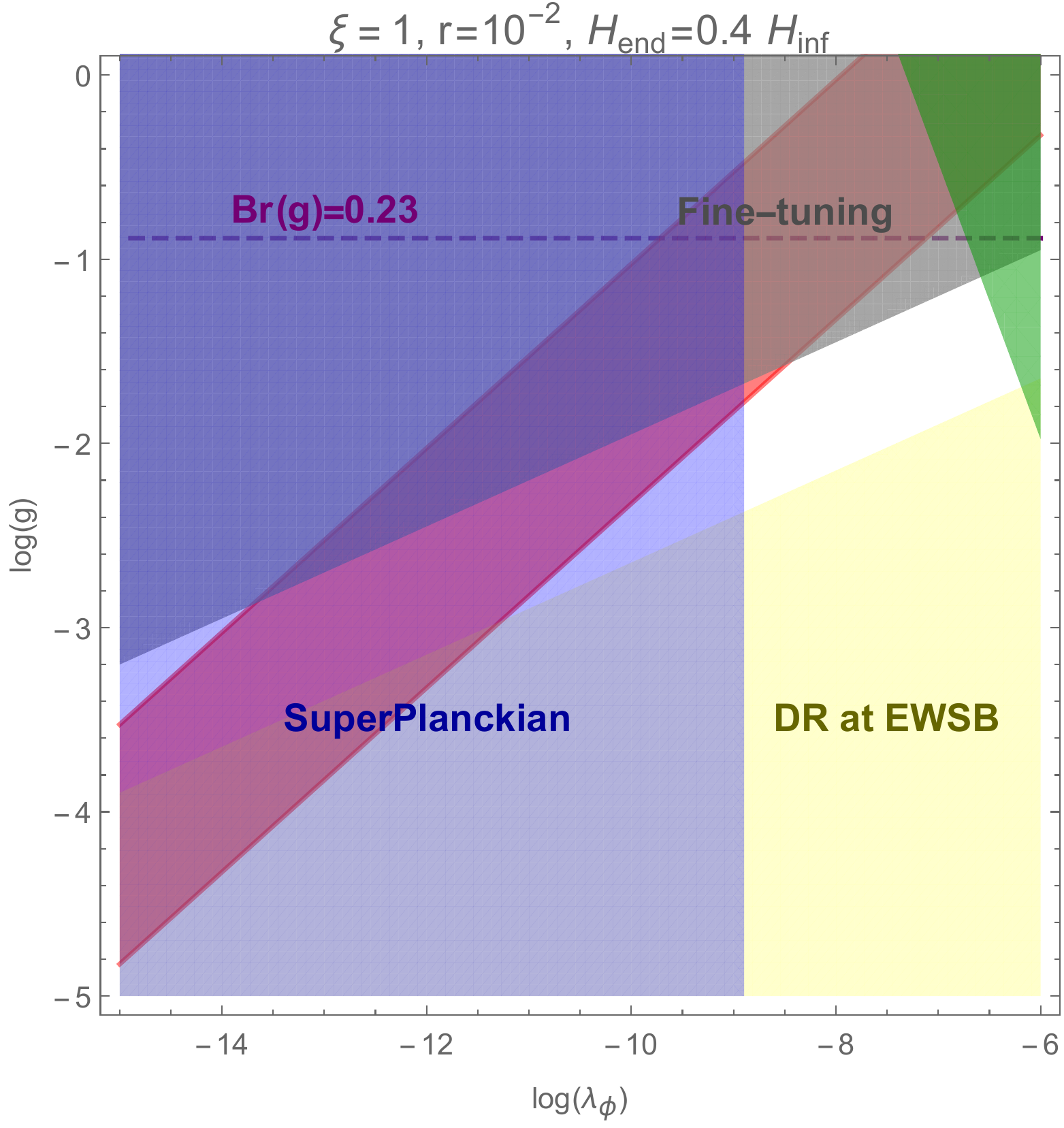}
  \end{minipage}

  \par\end{centering}
  \caption{Regions in the $\left(\lambda_{\phi},g\right)$ plane where
the constraints in Eqs. (\ref{condition: phi is not inflaton})
- (\ref{radiative corrections}) and Eq. (\ref{upper bound g reheating after}) are satisfied, for $r=10^{-2}$ and $\xi=0.1,1$. The red band encompasses the values of $g$ and corresponding $\lambda_{\phi}$ that can account for the present dark matter abundance, if $\phi$ makes up all the dark matter, for $10\,\mathrm{MeV}<T_{R}<80\,\mathrm{GeV}$ (the upper line in the red band corresponds to $T_{R}=10\,\mathrm{MeV}$ and the lower line corresponds to $T_{R}=80\,\mathrm{GeV}$). The excluded regions correspond to fine-tuned models (dark gray), super-Planckian dark scalar vevs during inflation, i.e, $\phi_{inf}>M_P/\sqrt{\xi}$, (blue) and scenarios where the dark scalar behaves as dark radiation and not as dark matter at or after EWSB (yellow). The dashed purple line
yields the current experimental limit on the branching ratio of the Higgs invisible decays, $\mathrm{Br}_{inv}\lesssim0.23$ and the green region in the upper right corner corresponds to scenarios for which the condensate evaporates.} 
\label{After r2}
\end{figure}

In this section we summarize our results, taking into account all the different constraints analyzed earlier. We present the results for the regions in the $\left(\lambda_{\phi},g\right)$ plane where all model constraints are satisfied, namely  Eqs. (\ref{condition: phi is not inflaton})
- (\ref{radiative corrections}) and (\ref{upper bound g reheating after}).  We choose to represent the
results for values of the tensor-to-scalar ratio $r=10^{-2}$ and nonminimal coupling $\xi=0.1,1$, as illustrated  in Fig. \ref{After r2}.

In Fig. \ref{After r2}, we can see that there is a window where our model can explain all of the present dark matter abundance, for dark scalar masses larger than the ones we have obtained in previous Higgs-portal scenarios with an oscillating scalar field, an underlying scale invariance and standard Cosmology (i.e, no early matter era) \cite{Bertolami:2016ywc, Cosme:2017cxk, Cosme:2018nly}. In Ref. \cite{Bertolami:2016ywc} we have studied
the case where the oscillating scalar field dark matter has negligible
self-interactions. We have found that the range of masses of the dark
scalar consistent with cosmological constraints is $m_{\phi}\gtrsim10^{-5}-10^{-6}$
eV. In particular, for not too suppressed tensor-to-scalar ratio,
$r$, this model predicts Higgs portal couplings of the order of $g\sim10^{-16}$.
In turn, Refs. \cite{Cosme:2017cxk, Cosme:2018nly} consider an oscillating
scalar field dark matter with non-negligible self-interactions. In
the case it acquires a vev after Electroweak symmetry breaking, it
may decay into photon pairs, with a mean lifetime larger than the
age of the Universe, predicting the observed galactic and extragalactic
3.5 keV X-ray line. In this scenario, the dark matter candidate has
a mass of about 7 keV, corresponding to $g\sim10^{-8}$. In the current
paper, for instance, we can see that $g\sim10^{-2}$ is allowed, which corresponds to $m_{\phi}\sim1$ GeV. We may conclude that an early matter era precludes sub-GeV dark scalar masses, mainly since these would lead to super-Planckian dark scalar values during inflation that could affect the latter's dynamics.

In fact, it is possible to get an analytic expression for the window of possible values for g and $\lambda_{\phi}$. Hence, since the dark scalar cannot affect inflation [Eq. (\ref{condition: phi is not inflaton})], and using the relation between the Higgs-portal coupling and the dark scalar's quartic coupling [Eq. (\ref{relation between g and lambda})], the constraint on $g$ becomes
\begin{equation}
g>5\times10^{-2}\left(\frac{T_{R}}{10\,\mathrm{GeV}}\right)^{-\frac{1}{3}}\left(\frac{r}{0.01}\right)^{\frac{1}{3}}\xi^{\frac{1}{2}}\left(\frac{H_{end}/H_{inf}}{0.4}\right)^{\frac{2}{3}},
\end{equation}
which translates into
\begin{equation}
m_{\phi}>8\,\left(\frac{T_{R}}{10\,\mathrm{GeV}}\right)^{-\frac{1}{3}}\left(\frac{r}{0.01}\right)^{\frac{1}{3}}\xi^{\frac{1}{2}}\left(\frac{H_{end}/H_{inf}}{0.4}\right)^{\frac{2}{3}}\mathrm{GeV}~.\label{up bound mass not inflaton}
\end{equation}
In turn, requiring that the field behaves like CDM at EWSB (Eq. (\ref{condition: CDM at EWPT})), and using the relation between couplings (Eq. (\ref{relation between g and lambda})), we find a lower bound on $g$: 
\begin{equation}
g>\frac{\sqrt{2\,\lambda_{h}}}{9\times10^{2}}\left(\frac{T_{R}}{10\,\mathrm{GeV}}\right)^{\frac{1}{3}}\,\left(\frac{r}{0.01}\right)^{\frac{1}{6}}\,\xi^{\frac{1}{2}}\,\left(\frac{H_{end}/H_{inf}}{0.4}\right)^{-\frac{2}{3}},
\end{equation}
and, consequently, a lower bound on the mass 
\begin{equation}
m_{\phi}\gtrsim0.1\,\left(\frac{T_{R}}{10\,\mathrm{GeV}}\right)^{\frac{1}{3}}\left(\frac{r}{0.01}\right)^{\frac{1}{6}}\xi^{\frac{1}{2}}\left(\frac{H_{end}/H_{inf}}{0.4}\right)^{-\frac{2}{3}}\mathrm{\,GeV}~.\label{lower bound mass DR after .}
\end{equation}
The no fine-tuning constraint allows only Higgs portal couplings below the following threshold:  
\begin{equation}
g<\frac{\sqrt{16\pi^{2}}}{9\times10^{2}}\,\left(\frac{T_{R}}{10\,\mathrm{GeV}}\right)^{\frac{1}{3}}\left(\frac{r}{0.01}\right)^{\frac{1}{6}}\xi^{\frac{1}{2}}\left(\frac{H_{end}/H_{inf}}{0.4}\right)^{-\frac{2}{3}},
\end{equation}
thus imposing an upper bound on the mass: 
\begin{equation}
m_{\phi}\lesssim2\,\left(\frac{T_{R}}{10\,\mathrm{GeV}}\right)^{\frac{1}{3}}\,\left(\frac{r}{0.01}\right)^{\frac{1}{6}}\,\xi^{\frac{1}{2}}\,\left(\frac{H_{end}/H_{inf}}{0.4}\right)^{-\frac{2}{3}}\mathrm{\,GeV}.\label{upper bound mass fine tuning after}
\end{equation}

\begin{figure}[htbp]
\centering
  \begin{minipage}[b]{0.4\textwidth}
  \includegraphics[width=1\textwidth]{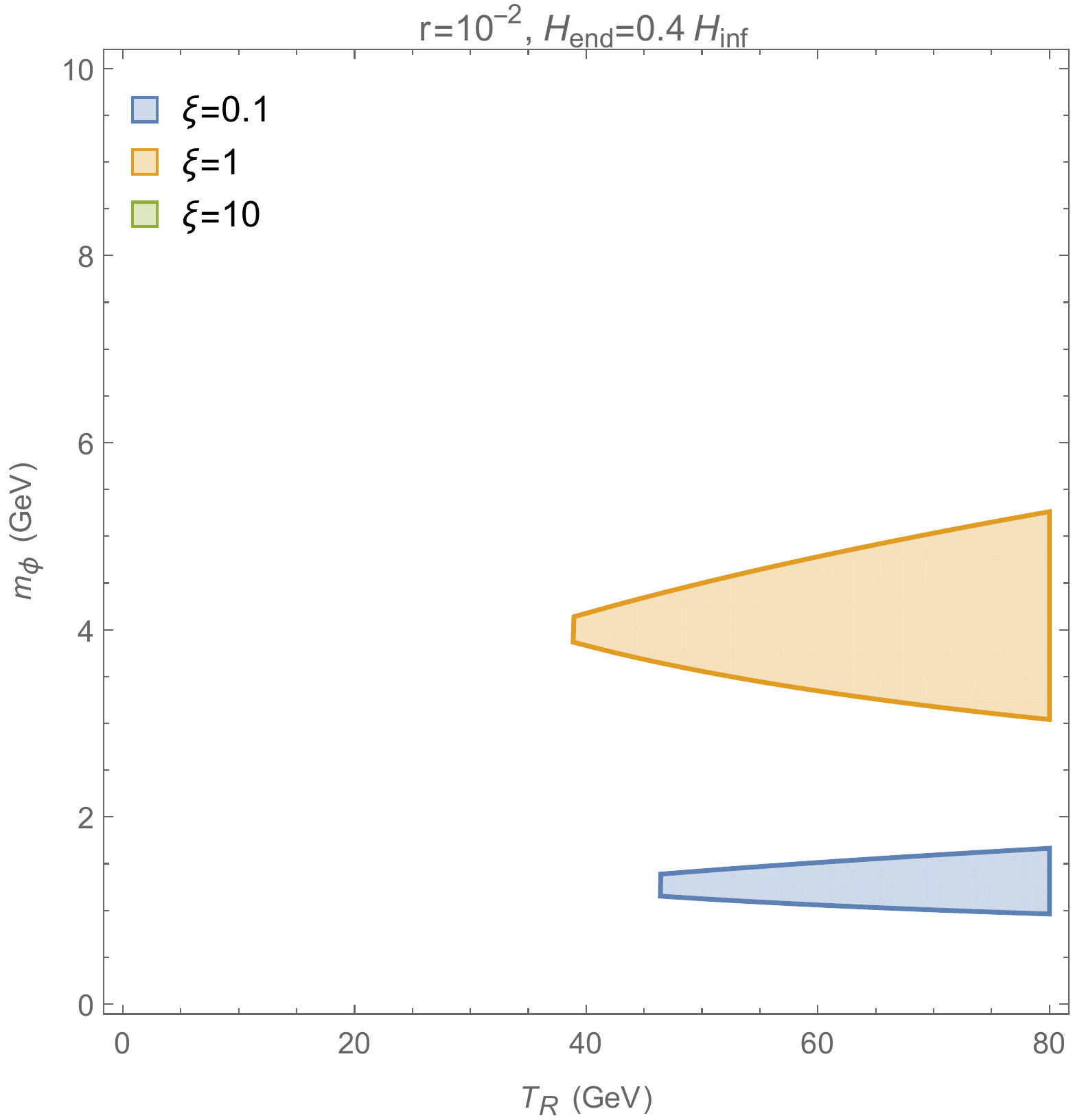}
  \end{minipage}
    \begin{minipage}[b]{0.4\textwidth}
  \includegraphics[width=1\textwidth]{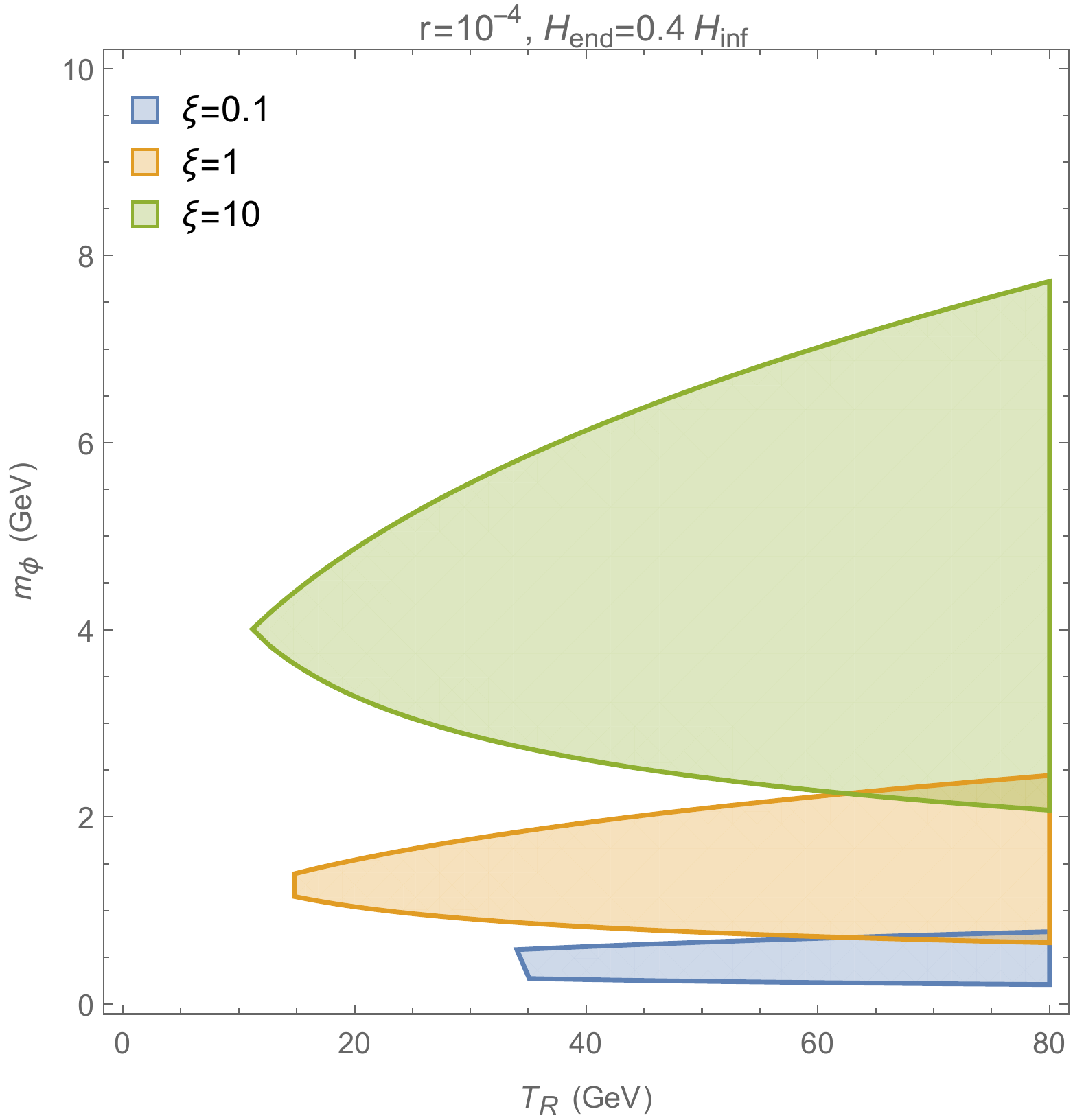}
  \end{minipage}
  \caption{Allowed values for the dark scalar mass as a function of the reheating temperature, for $10\,\mathrm{MeV}<T_{R}<80\,\mathrm{GeV}$ and considering different values for the nonminimal coupling to curvature $\xi$  and tensor-to-scalar ratio $r$.}
\label{After mass Trh}
\end{figure}

Taking into account all these restrictions, along with the bound coming from avoiding condensate evaporation, Eq. (\ref{upper bound g final reheating after}) and the LHC bound on the Higgs invisible partial decay width, Eq. (\ref{branch bound}), we may alternatively plot the allowed parametric regions in the ($m_\phi, T_R$) plane for different values of the nonminimal coupling to gravity and tensor-to-scalar ratio, as illustrated in Fig. \ref{After mass Trh}.

From Fig. \ref{After mass Trh}  we may conclude that our model predicts masses for the dark scalar in the few GeV range, depending on the values of the tensor-to-scalar ratio and nonminimal coupling chosen. These may be within the reach of the LHC or its successors in the near future, since for instance $\mathrm{Br}_{inv}\simeq2\times10^{-3}$ for $m_{\phi}\simeq6$ GeV, which is not too far from the current experimental limit (Eq. (\ref{branching ratio})). Notice, however, that large values of the nonminimal coupling to gravity, permitting heavier dark scalars,  are only allowed for lower values of $r$, i.e.~in scenarios with a low inflationary scale.


\section{Conclusions}  \label{discussion 3}

In this work, we have analyzed the possibility of an oscillating scalar field, accounting for all the dark matter in the Universe, driving a nonthermal spontaneous breaking of the electroweak symmetry. The dark scalar is coupled to the Higgs field through a standard ``Higgs-portal" biquadratic term, has no bare mass terms due to an underlying scale invariance of the theory, and has a negative nonminimal coupling to curvature. The latter, in particular, allows the dark scalar to develop a large expectation value during inflation. This holds the Higgs field at the origin both during and after inflation, until the dark scalar's oscillation amplitude drops below a critical value at which EWSB takes place. This prevents, in particular, the Higgs field from falling into the putative global minimum at large field values during inflation, ensuring at least the metastability of the electroweak vacuum.

The proposed scenario assumes a late decay of the inflaton field, such that reheating does not restore the electroweak symmetry, while the reheating temperature is still large enough to ensure a successful primordial nucleosynthesis \cite{footnote2}. Therefore, the Universe is still dominated by the inflaton field for a parametrically long period after inflation, while it oscillates about the minimum of its potential and behaves as a pressureless fluid. In fact, we have shown that consistent scenarios require reheating to occur only after EWSB, such that the latter occurs in the inflaton matter-dominated epoch essentially in vacuum.

Compared to other scenarios of scalar field dark matter where the Higgs is the only source of mass for the dark scalar field  \cite{Bertolami:2016ywc, Cosme:2017cxk, Cosme:2018nly}, we have shown that this allows for larger Higgs-portal couplings and hence dark scalar masses, since there are no thermalized particles in the Universe that could lead to an efficient evaporation of the scalar condensate until EWSB takes place. The dark scalar's oscillations, while it behaves as dark radiation, could themselves lead to particle production, but this can either be kinematically blocked in the case of Higgs production or made less efficient by the faster expansion of the Universe in a matter-dominated regime, as compared to the standard radiation epoch.

Overall, we have concluded that consistent scenarios where the dark scalar (1) does not affect the inflationary dynamics, (2) has technically natural values for its self-coupling (i.e.~requiring no fine tuning), and (3) starts behaving as cold dark matter after it breaks the electroweak symmetry, require dark scalar masses in the few GeV range, unless inflation occurs much below the grand unification scale. This looks promising from the experimental perspective, since it allows for Higgs invisible branching ratios $\lesssim 10^{-3}$, which may be within the reach of colliders in a hopefully not too distant future.

We thus reply ``Yes, it can" to the fundamental question posed in this work and hope that testing this idea may shed a new light on the nature of dark matter and on its role in the cosmic history.

\vspace{1.cm} 


\begin{acknowledgments}

C.\,C. is supported by the Arthur B. McDonald Institute (Canadian Astroparticle Physics Research Institute) and was supported by the Funda\c{c}\~{a}o para a Ci\^{e}ncia e Tecnologia (FCT) Grant No. PD/BD/ 114453/2016 and by the CFP Project No.~UID/FIS/04650/2013. J.\,G.\,R. is supported by the FCT Grant No.~IF/01597/2015, the CFisUC strategic Project No.~UID/FIS/04564/2019 and partially by the FCT Project No. PTDC/FIS-OUT/28407/2017 and ENGAGE SKA (POCI-01-0145-FEDER-022217). 

 \vfill
\end{acknowledgments}

\end{document}